\documentclass[]{zakoproc}
\usepackage{graphicx}
\usepackage{wrapfig}
\begin{document}

\title{Laboratory experiments and simulations on jets}
\author{Mart\'{\i}n Huarte-Espinosa\footnote{martinhe@pas.rochester.edu},
Adam Frank and Eric G. Blackman}
\institute{Department of Physics and Astronomy, University of Rochester,
600 Wilson Boulevard, Rochester, NY, 14627-0171}
\markboth{M. Huarte-Espinosa}{Laboratory experiments, and simulations, on jets}

\maketitle

\begin{abstract}
Astrophysical jets have been studied with observations, theoretical
models and numerical simulations for decades. Recently, supersonic
magnetized jets have been formed in laboratory experiments of
high-energy density plasmas. I will review these studies and discuss
the experimental setup that has been used to form millimeter-scale
jets  driven by strong toroidal magnetic fields in a MAGPIE
generator. The  physical conditions of these experiments are
such that they can be scaled to astrophysical scenarios. These
laboratory jets  provide insights on the underlying physics of magnetic tower
jets and help constrain some models of
astrophysical jets. In this context, we also discuss the connection between the
laboratory jets and recent 3D-MHD numerical simulations of
Poynting flux dominated jets. The simulations allow us to
investigate the effects of thermal
energy losses and base rotation on the growth rate of kink mode
perturbations, and to compare the evolution of PFD
jets with a hydrodynamic counterpart of the same energy flux.
\end{abstract}

\section{Introduction} 
\vskip-.3cm

Jets are observed in Young Stellar Objects, post-AGB stars, X-ray
binaries, radio galaxies and other astrophysical objects.  Models
suggest that jets are launched and collimated inside their ``central
engine'' by a symbiosis of accretion, rotation and magnetic mechanisms
(Pudritz et al.~\cite{pudritz}). The engines cannot be directly
observed however, because telescopes have insufficient resolution.
Recently, laboratory astrophysics experiments have  
provided scale models of the launch and propagation of magnetized
jets with dimensionless parameters relevant for astrophysical systems (Lebedev
et~al.~\cite{leb5}; Ciardi et~al.~\cite{ciardi9}; Suzuki-Vidal
et~al.~\cite{suzuki}). These studies, when combined with numerical
simulations (Ciardi et~al.~\cite{ciardi7}; 
Huarte-Espinosa et~al.~\cite{we}),  can help to resolve unanswered
questions in jet physics. A fundamental distinction can be made between the physics
of jet launch and that of jet propagation far from the engine. Since
we cannot resolve the former observationally, it is important to
identify distinct features of jets in the asymptotic propagation
regime that can distinguish different engine paradigms. While both
simulations and experiments now consistently reveal the promise,
if not essentiality, of dynamically significant magnetic fields for
jet launch, the correlation between the initial jet magnetic
configuration and the stability of the flow far from the launching
region is unclear. The importance of the magnetic field relative
to the flows' kinetic energy divides jets into (i) Poynting flux
dominated (PFD; Shibata \& Uchida~\cite{shibata}), in which magnetic
fields dominate the jet structure, (ii) magnetocentrifugal (Blandford
\& Payne~\cite{bland}), in which magnetic fields only dominate out
to the Alfv\'en radius. The observable differences between PFD and 
magnetocentrifugal jets are unclear, as are the effects that cooling and 
rotation have on PFD jets.

\section{Laboratory experiments}
\vskip-.3cm

Experiments and numerical modelling of astrophysically relevant
supersonic plasma jets  have been performed by a number of authors
using high intensity lasers (see e.g. Farley et al.~\cite{laser1}; 
Shigemori et al.~\cite{laser2}; Foster et~al.~\cite{laser3}; 
see also review in Blackman 2007).  
In this section however, we focus on  magnetized jet experiments 
which have dimensionless numbers (Reynolds, magnetic Reynolds and 
Peclet) in reasonably appropriate regimes for astrophysics  and have
used conical wire arrays and conducting foils on pulsed power
facilities (Lebedev et al. 2005; Ciardi et al. 2007,2009; Suzuki-Videl
et~al. 2010).

\subsection{Wire array, toroidal field}
\vskip-.3cm

Lebedev et~al.~(\cite{leb5}) applied a TW electrical pulse (1\,MA,
250\,ns) to an array of wires located inside a vacuum chamber.  The
set up consisted of a pair of concentric electrodes connected radially
by tungsten wires of 13\,$\mu$m in diameter. The current causes
ablation of the wires which results in the formation of a
background ambient plasma (Fig.~1a). 
This material is then pushed above the
wires by Lorentz forces, and resistive diffusion keeps the current
close to the wires. The current induces a toroidal magnetic field
which at this stage is confined below
the wires, around the central electrode. Then, after the complete ablation
of the wires near the central electrode, the current switches to
the plasma and creates a magnetic cavity (Fig.~1b) with a jet at its center.
The core of the jet is confined and accelerated upward by
the pressure of the toroidal field. The return current flows along
the walls of the magnetic cavity, which is in turn confined by both
the thermal pressure and the inertia of the ambient plasma. Next,
the magnetic cavity opens up, the jet becomes detached and it
propagates away from the source at a velocity of order 200\,km\,s$^{-1}$ 
(Fig.~1c).

\begin{figure}[hb]
\begin{center}
\includegraphics[width=\textwidth]{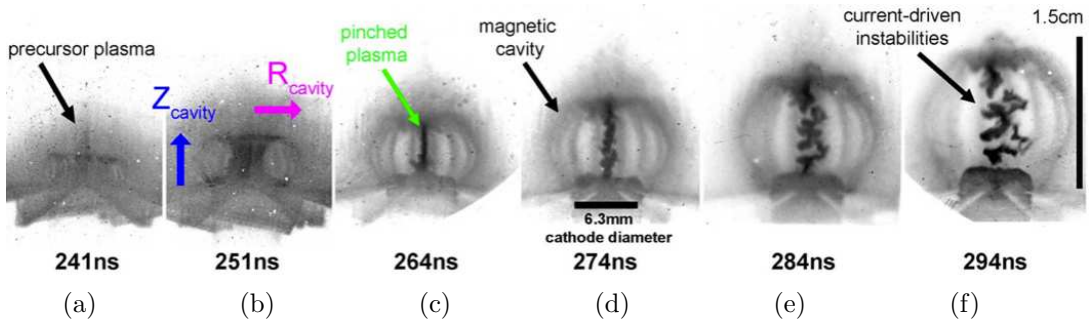}
~~~~~~~~
(a)
~~~~~~~~~~~~~~
(b)
~~~~~~~~~~~~~~
(c)
~~~~~~~~~~~~~~
(d)
~~~~~~~~~~~~~~
(e)
~~~~~~~~~~~~~~
(f)
~~~~~~
\end{center}
\vspace{-15pt}
\caption{Time sequence of soft X-rays images showing the formation
of the background plasma (a), the expansion of the magnetic cavity (b), the
launch of the jet (c), the development of instabilities in the central
jet column (d) and the fragmentation of the jet (e,f). 
The wire array shows in the bottom part of the figures.
Image taken from Suzuki-Vidal et~al.~(\cite{suzuki}).}
\end{figure}

Finally, instabilities, which resemble those of the kink mode ($m =\,$1), develop
within the body of the jet (Fig.~1d). The outflow is then fragmented into
well collimated structures with characteristic axial non-uniformities
(Fig.~1e,f; Lebedev et~al.~\cite{leb5}). 
We note that
a critical ingredient of these experiments is the significant thermal
energy loss of both the jet and the ambient plasmas. 
This is relevant because cooling plays a critical role in 
many astrophysical jet environments, e.g. YSOs.

The experiment was reproduced a few years later by
Ciardi et~al.~(\cite{ciardi7}) using 3-D non-ideal MHD simulations.
A numerical model was carefully  designed to simulate
the laboratory components (electrodes and wires) and all the plasma 
evolution phases. Good agreement was found between the simulations 
and the experiments (Ciardi et~al.~\cite{ciardi7}). The simulations showed
that during the final unstable phase of jet propagation, the magnetic
fields in the central jet adopted a twisted helical structure. This
confirmed that the jets are affected by normal mode, $m=\,$1, perturbations.

\subsection{Wire array, toroidal and axial fields}
\vskip-.3cm

A subsequent series of experiments were carried out by Suzuki-Vidal
et~al.~(\cite{suzuki}) who studied the effect of introducing an axial
magnetic field, $B_z$, into a radial wire array. Their experimental
configuration was the same as that of Lebedev et~al.~(\cite{leb5},
above), but they placed two solenoids below the wires, into the
path of the current. Suzuki-Vidal et~al.~(\cite{suzuki}) found that
the added $B_z$ affects the degree of compression of
the on-axis pinch of the jet. As they increased the magnitude of $B_z$,
the plasma column radially expanded after reaching a minimum radius.
They also saw that jets are more stable 
when $B_z$ is present than otherwise. We note that
this stability effect can be explained analytically (with perturbation theory
on a plasma column) and numerically (section~3) as a result of
magnetic pressure form $B_z$ acting against hoop
stress compression from the toroidal field component. 

\subsection{Conducting foil}
\vskip-.3cm

\begin{wrapfigure}[22]{r}{0.36\textwidth}
\vspace{-16pt}
\begin{center}
\includegraphics[width=.35\columnwidth]{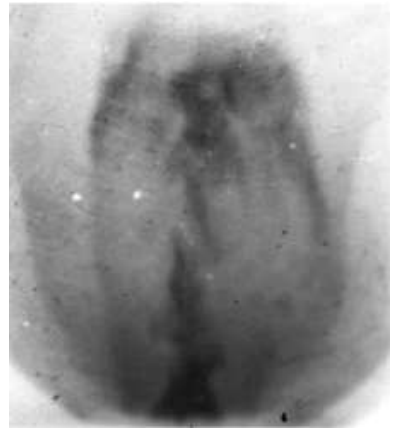}
\vspace{-15pt}
\end{center}
\caption{Filtered XUV emission images of one of the episodic jet
experiments. The clumpy jet at the center is
contained inside three nested magnetic bubbles which were formed 
by previously launched jets (Ciardi et~al.~\cite{ciardi9}).}
\end{wrapfigure}
Ciardi et~al.~(\cite{ciardi9}) extended the experiments of Lebedev
et~al.~(\cite{leb5}) by replacing the wire array (section~2.1) by a 6 $\mu$m
thick aluminum foil. Ciardi et~al.~(\cite{ciardi9}) found the following.
The conducting foil experiment produced very similar results than
the wire array one. However,
while the wire array experiment results in the launch of one jet only,
the conducting foil experiment results in a series of jets which
are launched sequentially (Fig.~2). The plasma flux caused by ablation of
the foil is much larger than that of the wires; the foil
provides an increased mass as a function of radius. Thus the current
gap produced by the magnetic cavity (section~2.1) is smaller in
the foil experiment than in the wire array one, and can be refilled
by the readily available plasma.
Closure of the gap, which does not happen in the wire array case,
allows the current to flow once again across the base of the magnetic
cavity, thus re-establishing the initial configuration.  Once the
magnetic pressure is large enough to break through this newly
deposited mass, a new jet/bubble ejection cycle begins (Ciardi
et~al.~\cite{ciardi9}). This constituted the first laboratory study
to address time-dependent episodic jet launch. 

\section{Simulations}
\vskip-.3cm
\subsection{Model}
\vskip-.3cm
We use the Adaptive Mesh Refinement code AstroBEAR2.0 (Cunningham
et~al.~\cite{bear}; Carroll-Nellenback et~al.~\cite{bear2}) to solve
the equations of radiative-MHD in 3D.  The grid represents
160$\times$160$\times$400\,AU divided into 64$\times$64$\times$80
cells plus 2 adaptive refinement levels.  We use periodic boundary
conditions at the four vertical faces of the domain, extrapolation
conditions at the top face and a combination of reflective and
inflow conditions at the bottom face. Initially, molecular gas
is static and has an ideal gas equation of state ($\gamma=\,$5$/$3),
a number density of 100\,cm\,s$^{-1}$ and a temperature of 10000\,K.
The magnetic field is helical, centrally localized and given by the
vector potential (in cylindrical coordinates) 

\begin{equation}
{\bf A}(r,z) = \left\{
   \begin{array}{c l}
          \frac{r}{4} (cos(2\,r) + 1)( cos(2\,z) + 1 ) \hat{\phi} + 
          \frac{\alpha}{8} (cos(2\,r) + 1)( cos(2\,z) + 1 ) \hat{k},
             & \mbox{for}~r,z < r_e; \\
          0, & \mbox{for}~r,z \ge r_e,
   \end{array} \right.
\label{apot} 
\end{equation}

\noindent where $r_e=\,$30\,AU and
$\alpha=\,$40 has units of length and determines the ratio of
toroidal to poloidal magnetic fluxes.
The magnetic pressure exceeds the thermal pressure
inside the magnetized region.

Source terms continually inject magnetic or kinetic
energy at cells $r,z<r_e$.  We carry out 4 simulations: an
adiabatic, a cooling, a rotating PFD jet, and  a hydrodynamical
jet. The latter is constructed  to have same time average propagation speed and energy
flux as the adiabatic PFD jet. The cooling tables of Dalgarno \&
McCray~(\cite{dm}) were used for the cooling case. For the rotating case
we allowed the gas and magnetic fields within $r,z<r_e$ to
move in Keplerian rotation.

\begin{wrapfigure}[21]{r}{0.66\textwidth}
\vspace{-42pt}
\begin{center}
\includegraphics[width=0.65\textwidth]{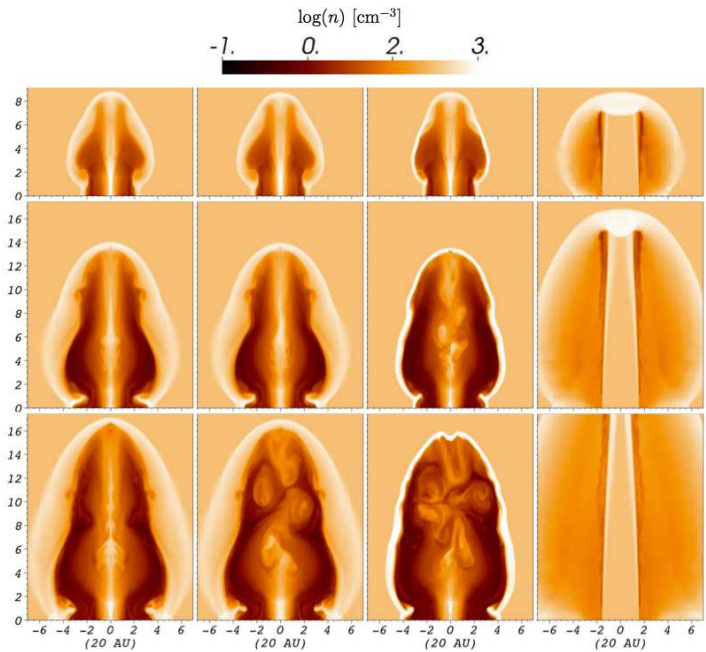} \\
\end{center}
\vspace{-20pt}
\caption{Logarithmic density maps of the adiabatic (1st column), rotating
(2nd column) and cooling (3rd column) PFD jets. Hydrodynamic jet 
(4th column). From top to bottom the time is 42, 84 and 118\,yr.}
\vspace{-10pt}
\end{wrapfigure}

\subsection{Results}
\vskip-.3cm

Magnetic pressure pushes  field lines and plasma upward, forming magnetic
cavities with low density (Fig.~3, all but right panel). 
The adiabatic case is the most stable.
PFD jets decelerate more quickly relative to our hydro jet. This results because the PFD and hydro jets have the  same injected energy flux, but the PFD case produces not only axial but radial expansion.
The pre-collimated hydro  can only expand via a much lower thermal pressure. Thus all of the energy flux in the hyrdo-case for our set up is more efficiently directed to axial mechanical power.  
In principle, our hydro case can emulate the asymptotic propagation regime of a jet that was magneto-centrifugally launched (e.g. Blackman 2007) which is distinct from a PFD jet.

The PFD jet cores are thin and unstable, whereas the hydro jet beam is thicker,
smoother and stable.   The PFD jets are sub-Alfv\'enic. Their cores
are confined by magnetic hoop stress, while their surrounding
cavities are collimated by external thermal pressure. PFD jets carry
high axial currents which return along their outer contact discontinuity. 

We see that the inner regions of the PFD jets, just outside the core, are low beta plasma columns in which 
the axial magnetic field dominates over the toroidal one, 
$|B_{\phi} / B_z | \ll 1$. The columns' instability condition is given by
\begin{equation}
   \left| \frac{B_{\phi}}{B_z} \right| > | (\beta_z - 1)k r_{jet}  |,
   \label{insta}
\end{equation}
\noindent where $\beta_z=2 \mu_0 P / B_z^2$, 
$\mu_0$ is the magnetic vacuum permeability,
$P$ is the plasma thermal pressure and $k^{-1}$ is the
characteristic wavelength of the current-driven perturbations. 
We find that the cooling jet shows $\beta_z \sim\,$1
from early in the simulation (Fig.~4), and thus  
does not have sufficient thermal energy to damp the
magnetic pressure kink perturbations. The latter grow
exponentially.

\begin{figure}[ht]
\begin{center}
\includegraphics[scale=.32,angle=90]{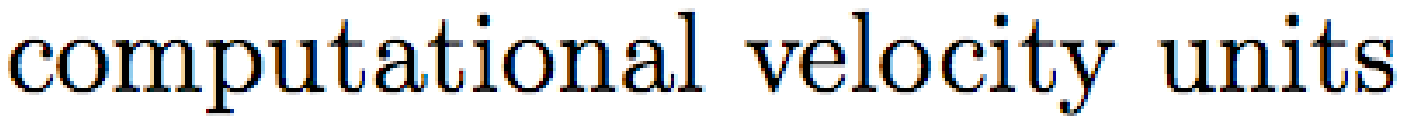} 
\includegraphics[width=.315\columnwidth,bb=70  150 560 665,clip=]{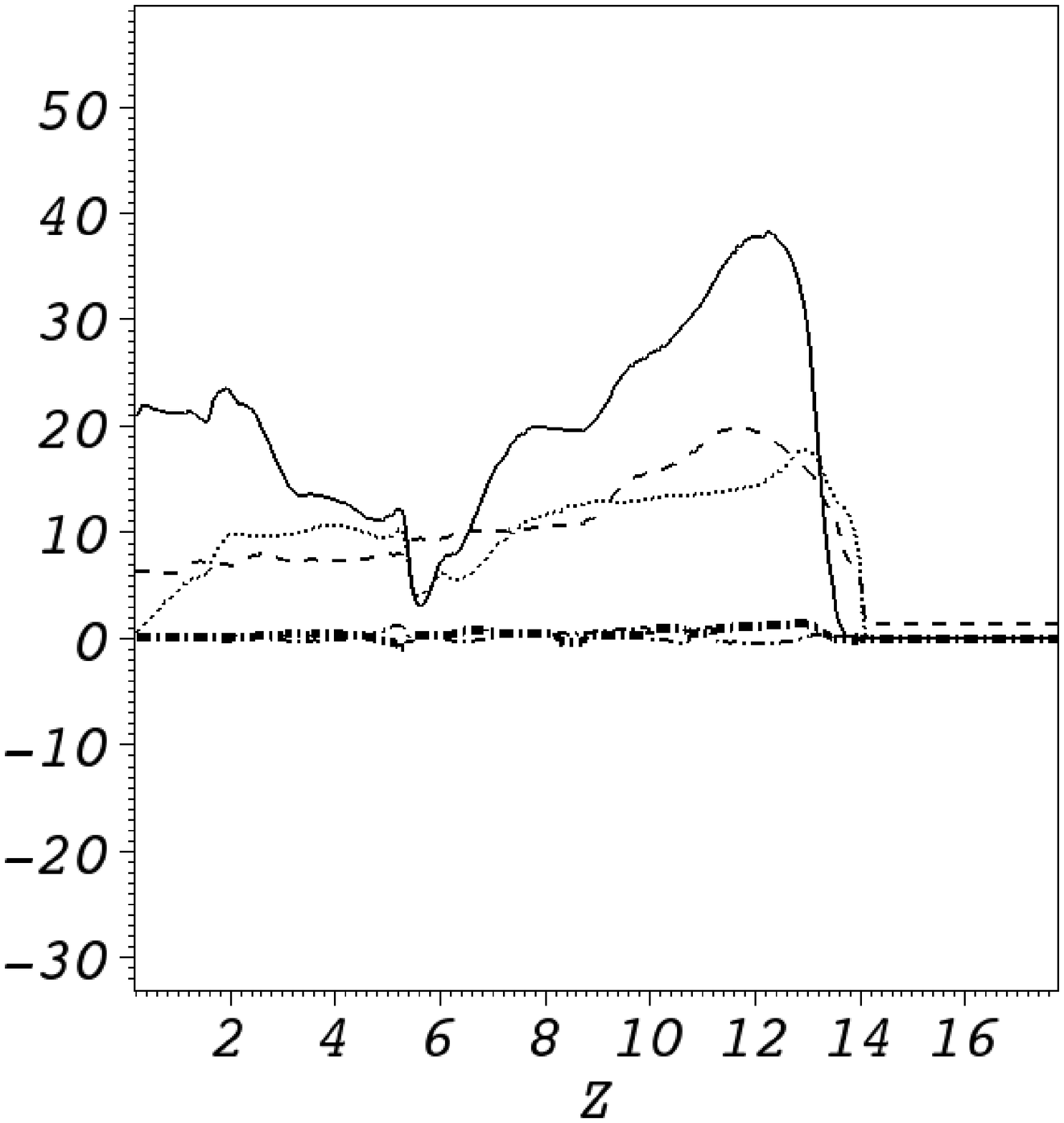} 
\hskip-.1cm
\includegraphics[width=.29\columnwidth,bb=125 150 560 645,clip=]{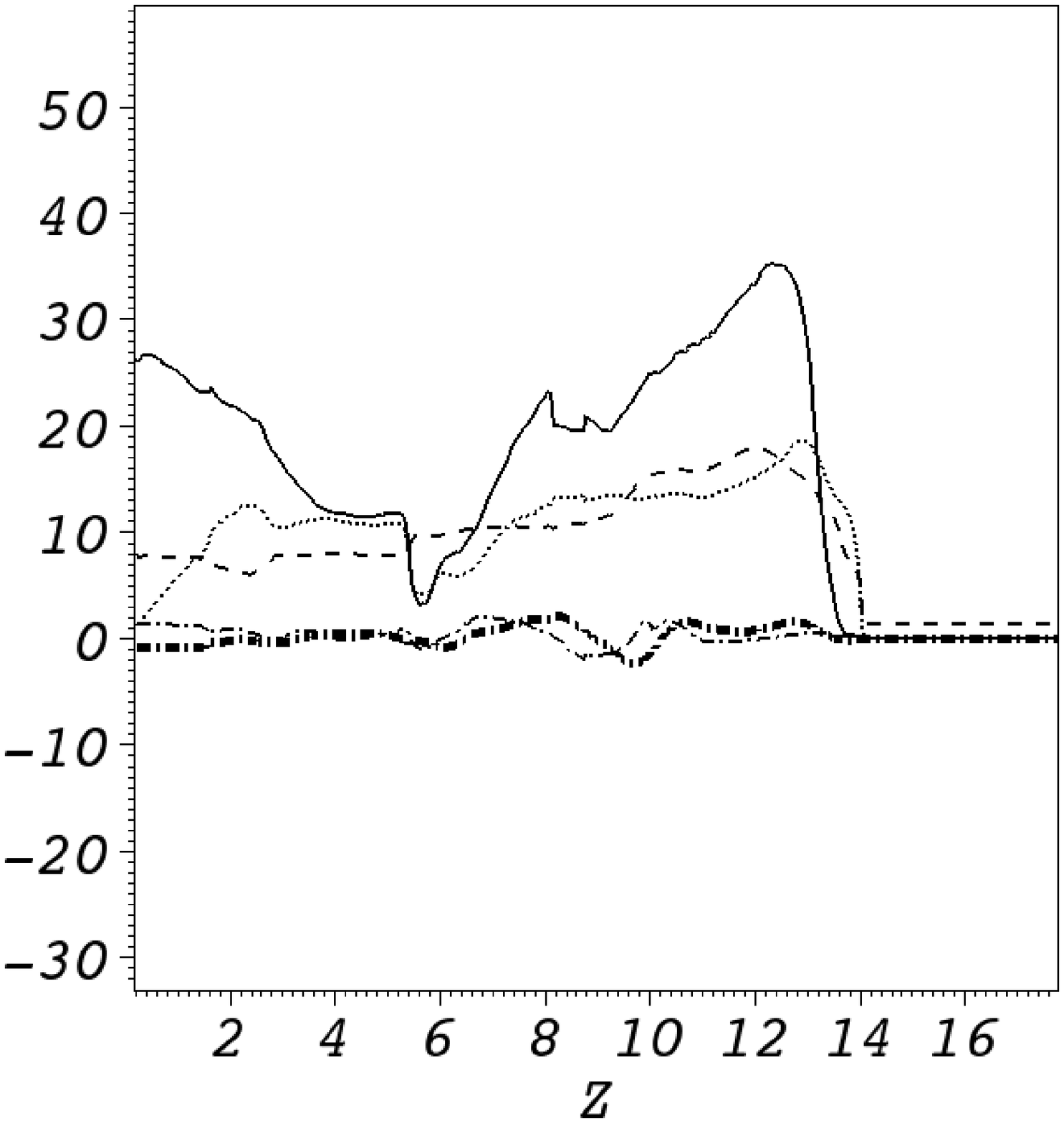} 
\hskip-.1cm
\includegraphics[width=.29\columnwidth,bb=125 150 560 645,clip=]{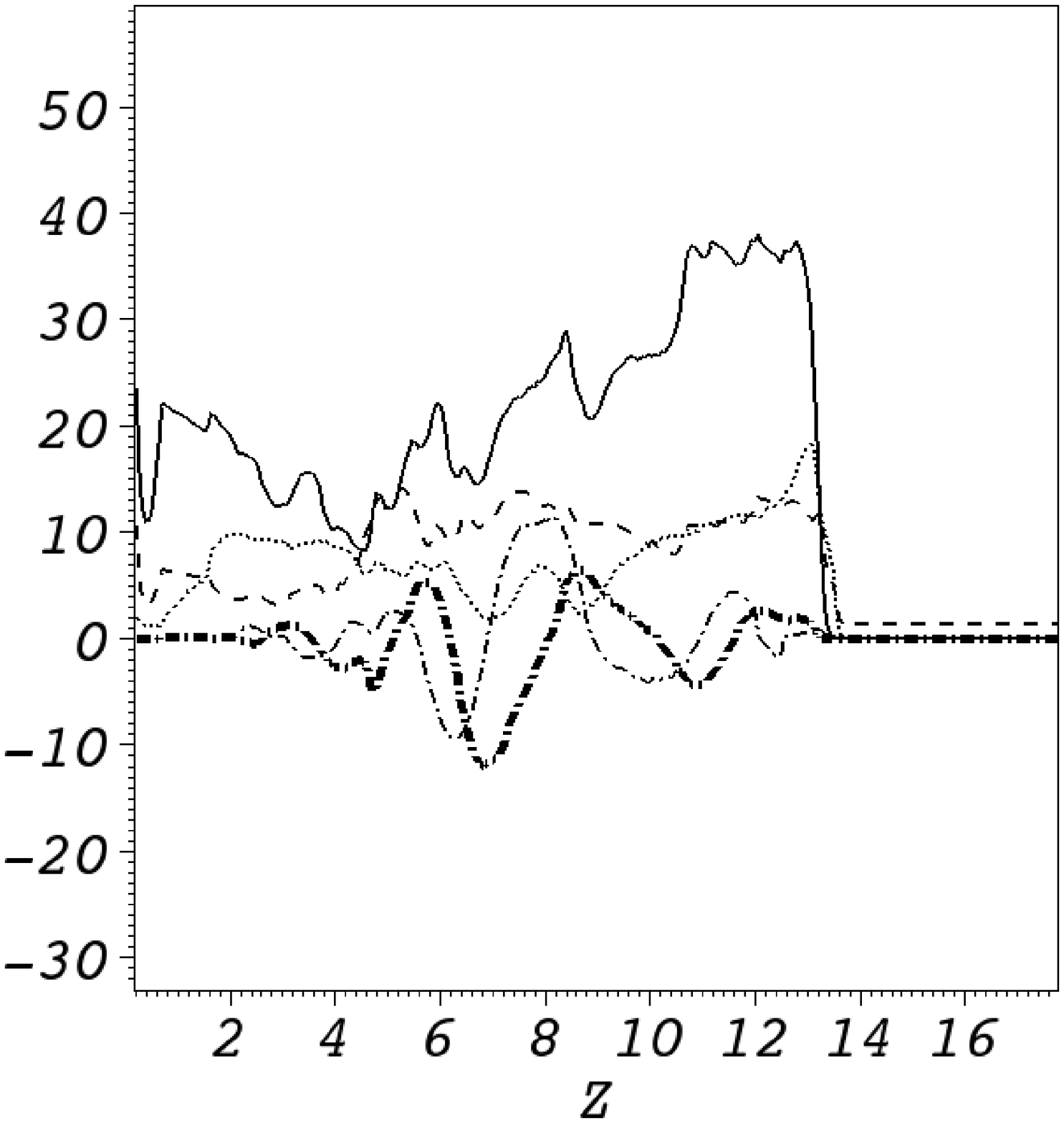} \\
Adiabatic
~~~~~~~~~~~~~~~~~~~~~~~~
Rotating
~~~~~~~~~~~~~~~~~~~~~~~~
Cooling
\end{center}
\vspace{-15pt}
\caption{Plasma velocities along the axes of the PFD jets after expanding for 84\,yr.
The solid, dashed, dotted, dot-dash and dot-dash thick
lines represent the Alfv\'en speed, the sound speed, $v_z, v_x$ and $v_y$,
respectively. Each velocity unit represents 9.1\,km\,s$^{-1}$.}
\end{figure}

A different path to instability operates  for the rotating (non-cooling) case.
For this case, rotation at the base of the jet causes a  slow amplification
of the toroidal magnetic field.  Hence the 
left hand side of equation~(\ref{insta}) increases slowly, and so
do the kink mode perturbations.  The rotating jet is not completely
destroyed by these perturbations, and their amplitude is about twice
the radius of the central jet (Fig~5), in agreement with the
Kruskal-Shafranov criterion (Kruskal et~al.~\cite{kruskal};
Shafranov~\cite{shafranov}). \\

In Figure~4 we show profiles of the relevant velocities of the PFD
jets along
the jet axis, as a function of cooling and rotation.
We also followed these in time.  During the stable propagation
phase, we find that the jets are mostly sub-Alfv\'enic and trans-sonic,
independent of cooling or rotation.  Fast-forward compressive MHD
(FF) and transmitted hydrodynamic shocks are evident in the ambient
medium, ahead of the jets.  The FF shocks steepen in time,
whereas the hydrodynamic shocks are quickly dissipated in the cooling
case.  In contrast, the adiabatic and rotating cases show regions
within the lower half of the jets, where the sound speed
is super-Alfv\'enic. Such regions are bounded by the reverse and
the forward slow-modes of compressive MHD waves, and characterized
by high thermal to magnetic pressure ratios.

\section{Conclusions}
\vskip-.3cm
\begin{wrapfigure}[29]{r}{0.69\textwidth}
\vspace{-39pt}
\begin{center}
Magnetic field strength [$\mu$G]\\
\includegraphics[width=.23\columnwidth,bb=5 605  375 670,clip=]{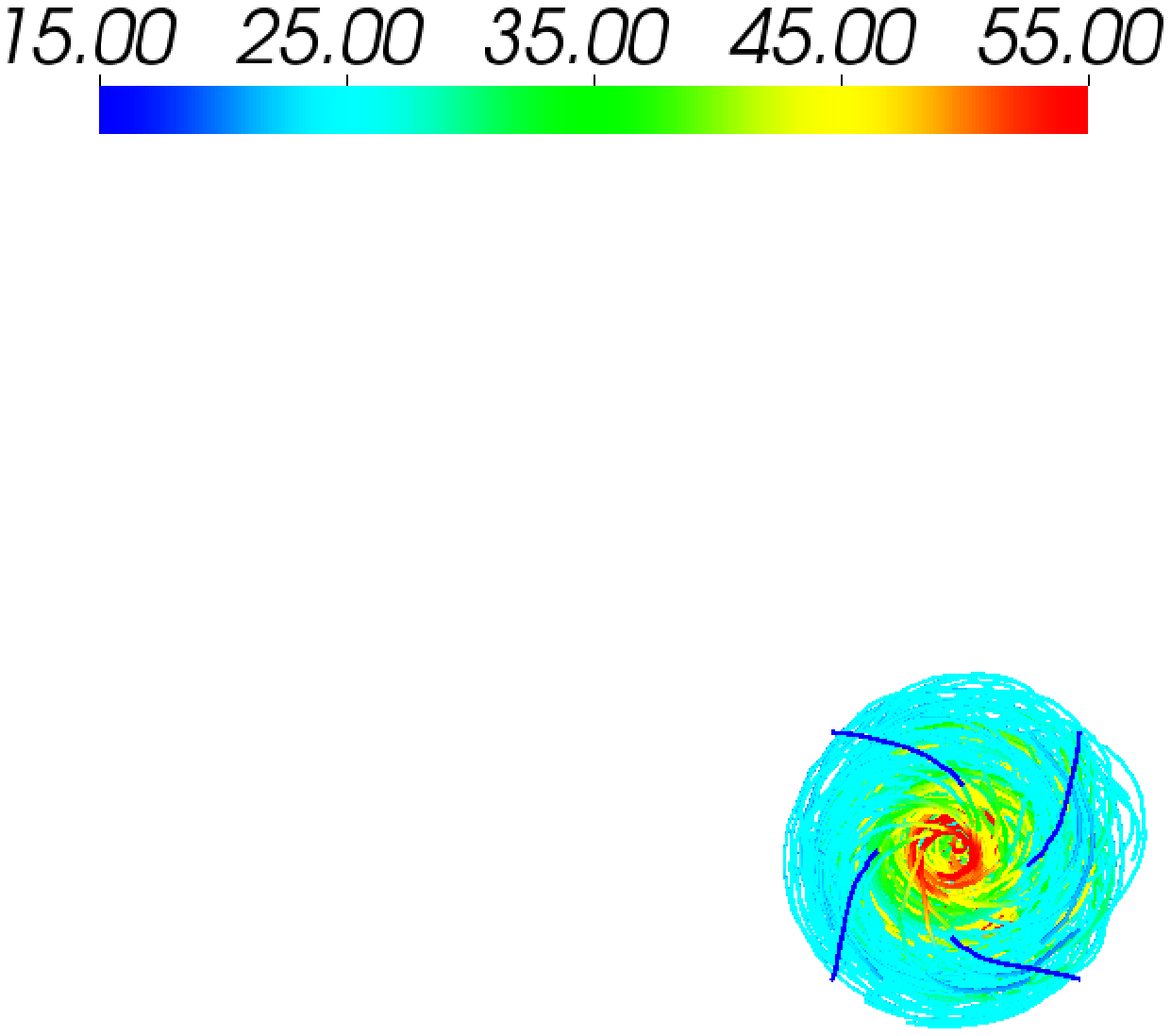} \\
\includegraphics[width=.22\columnwidth,bb=160 120  450 700,clip=]{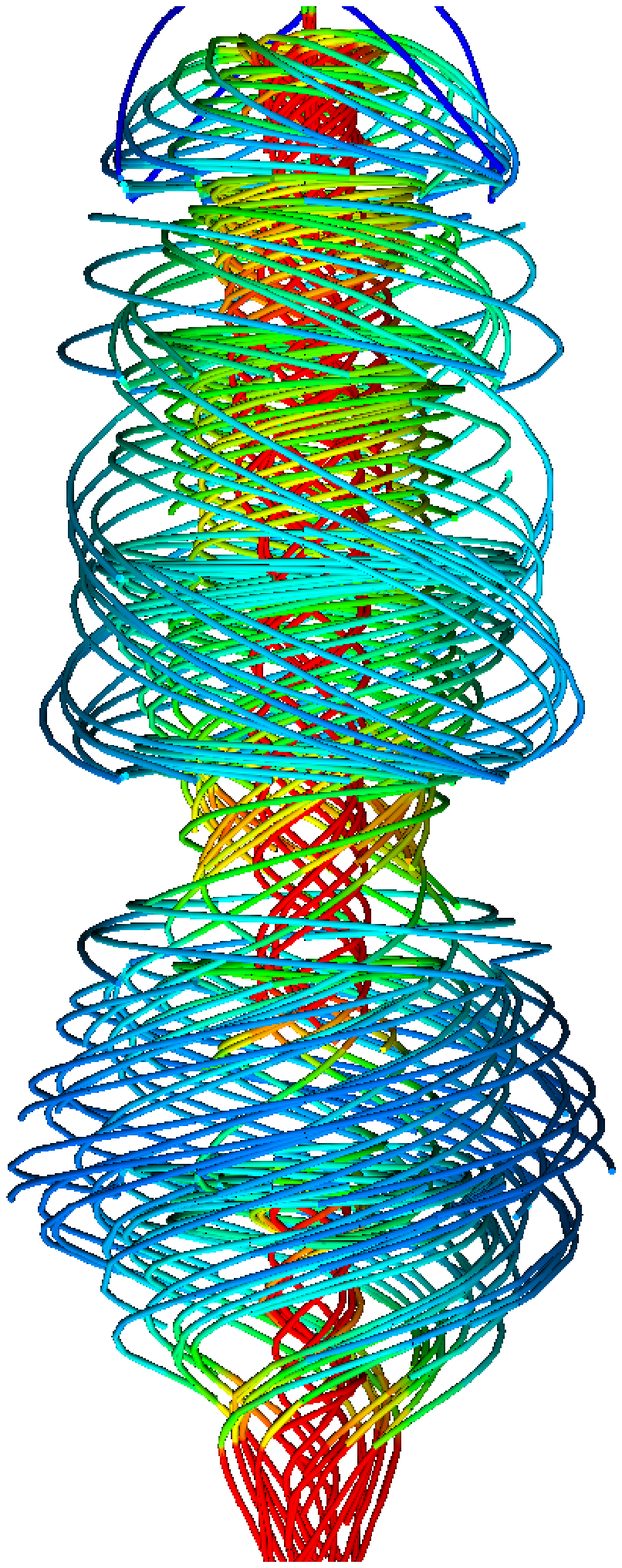}
\includegraphics[width=.22\columnwidth,bb=160 120  450 700,clip=]{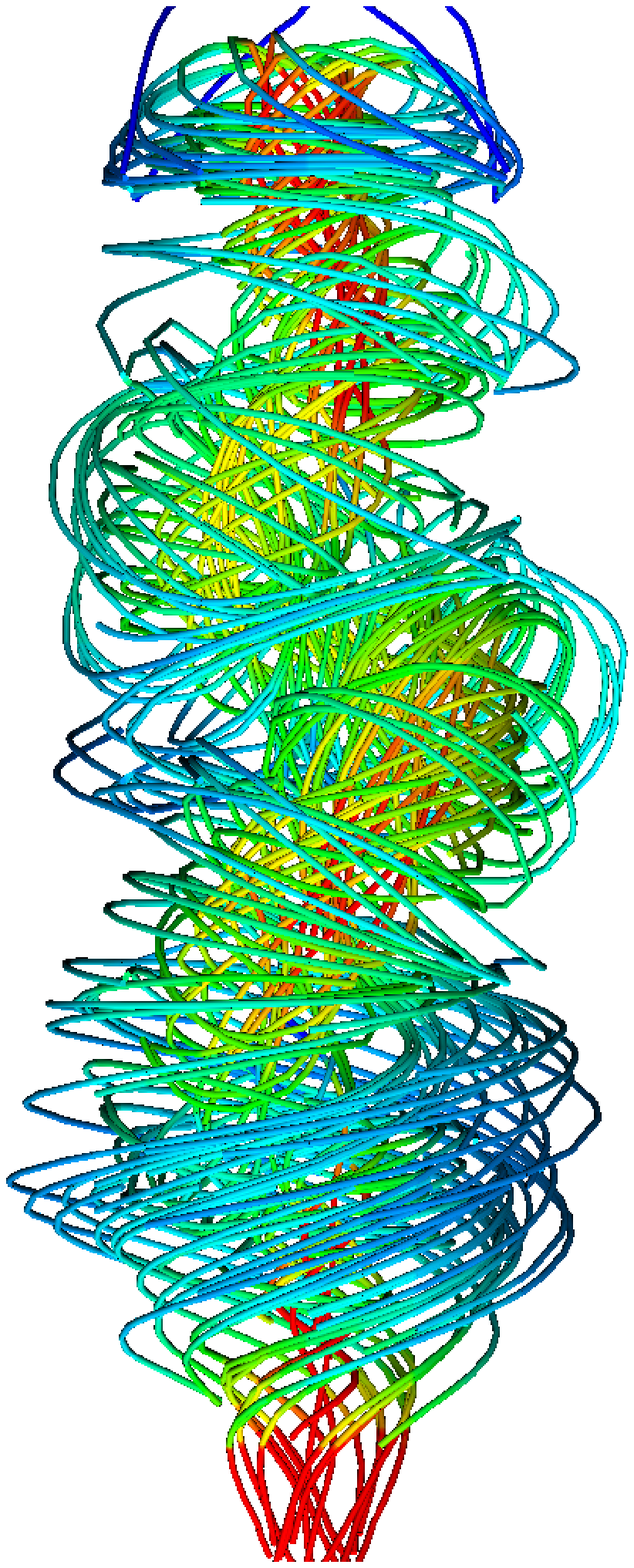}
\includegraphics[width=.22\columnwidth,bb=160 120  450 700,clip=]{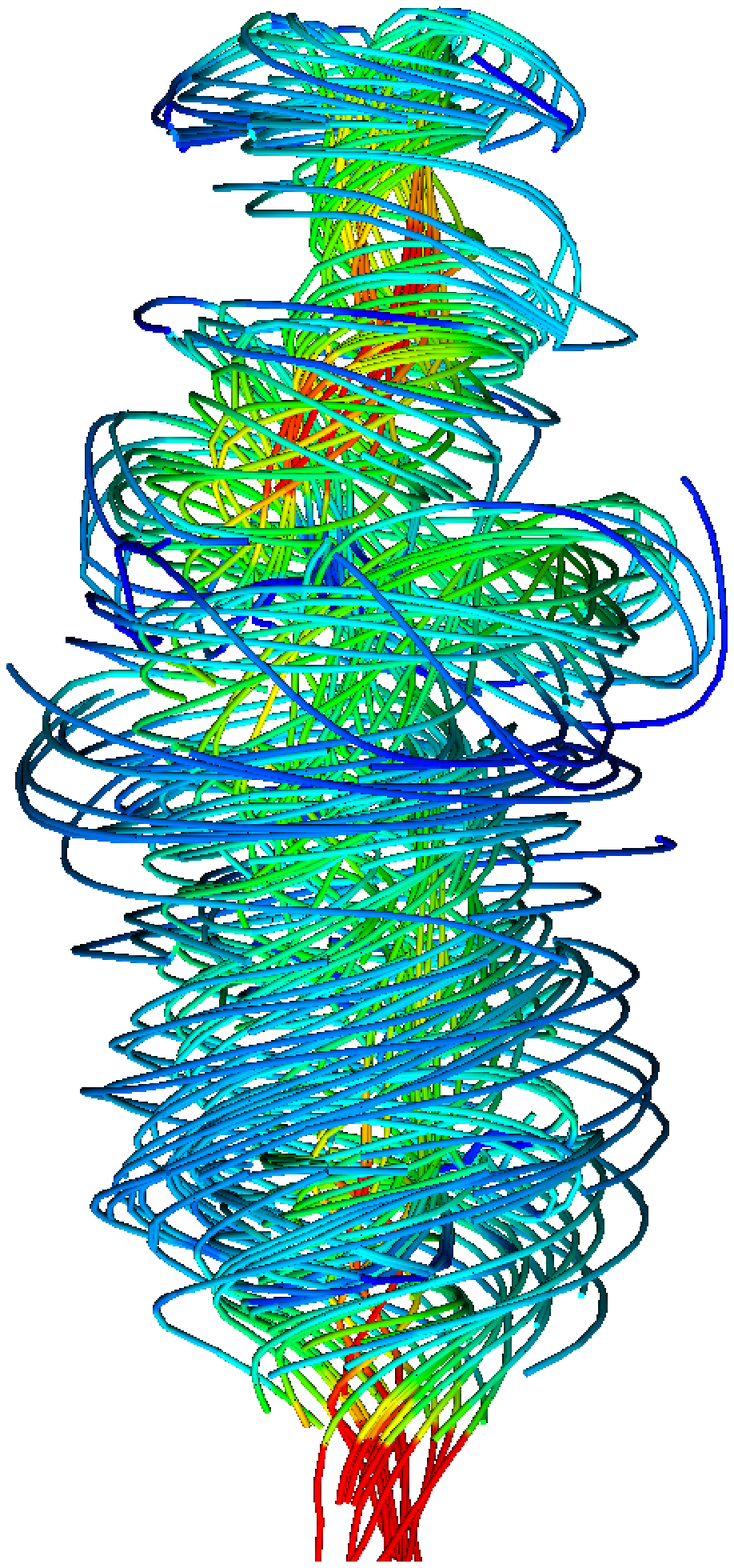} 
\includegraphics[width=.22\columnwidth]{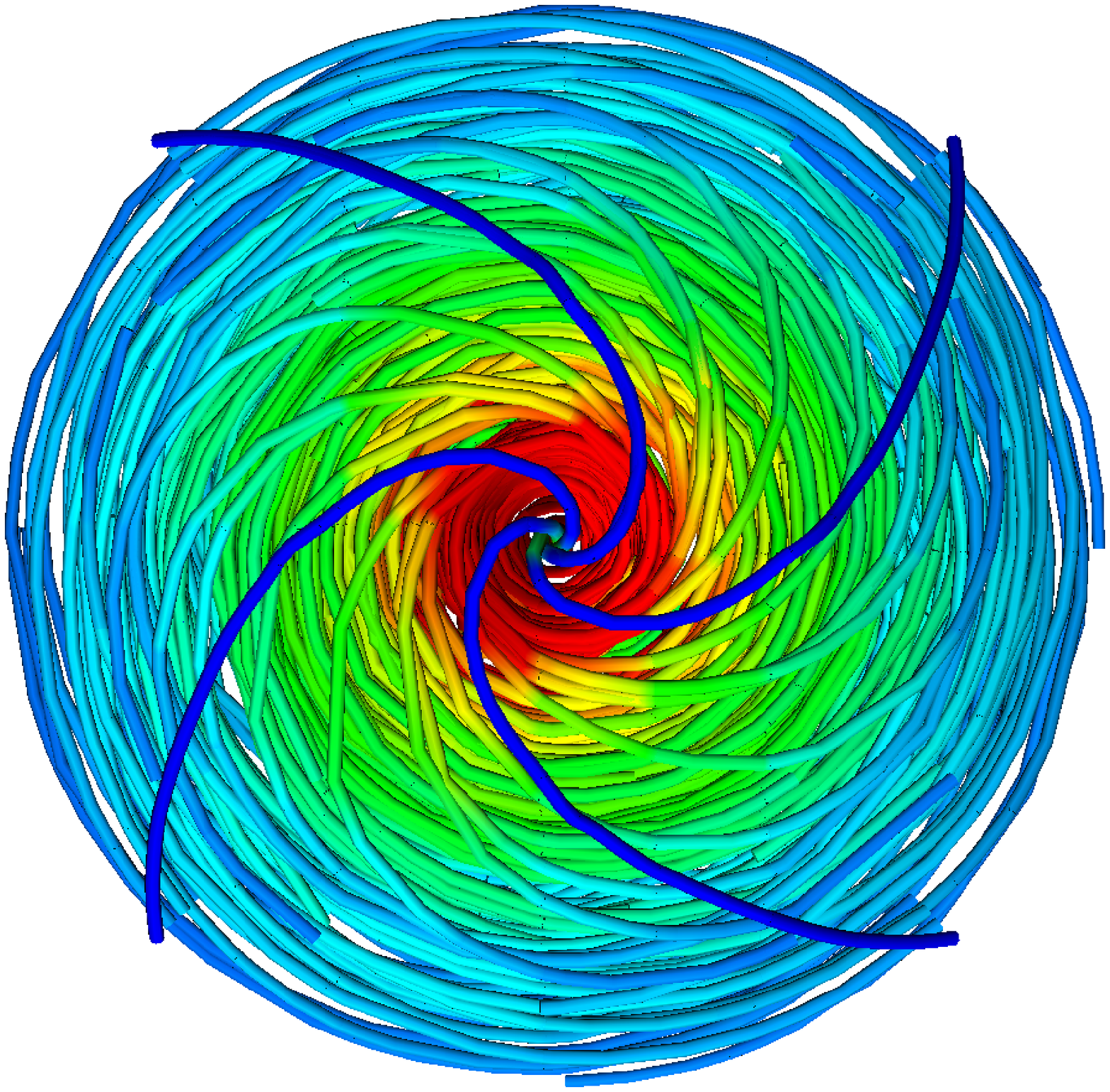}
\includegraphics[width=.22\columnwidth]{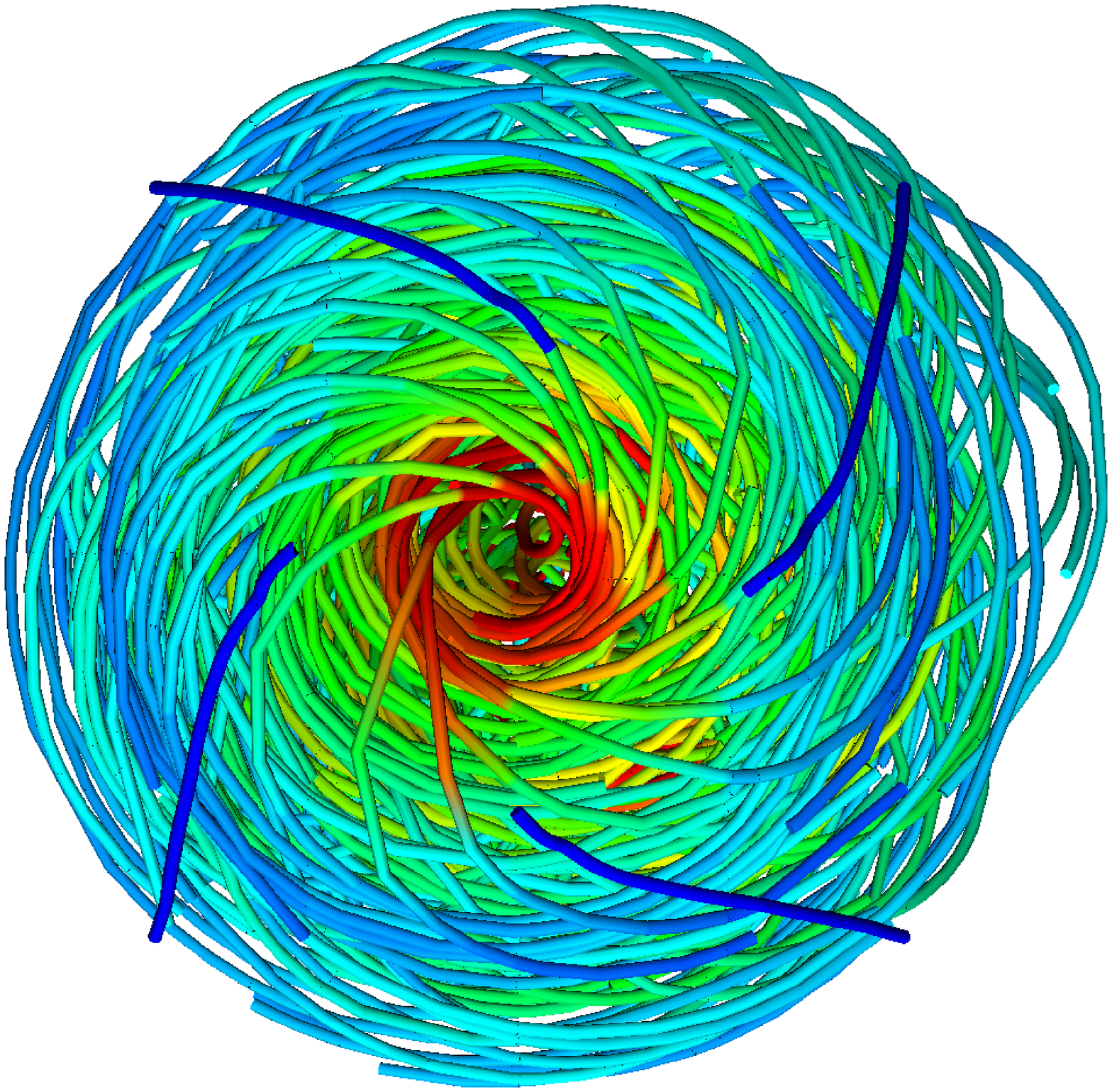}
\includegraphics[width=.22\columnwidth]{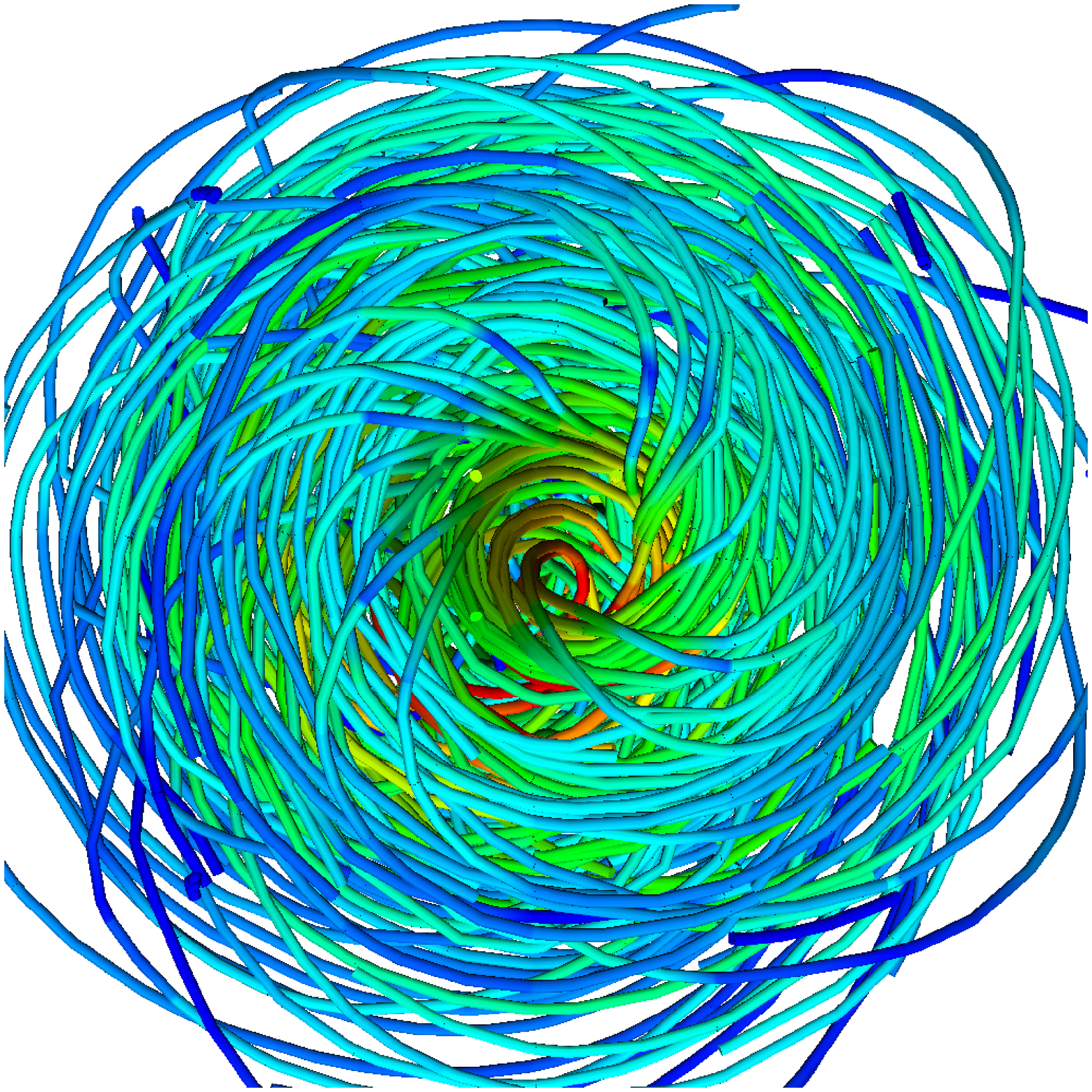}
\end{center}
\vspace{-10pt}
\caption{Central magnetic field lines at
$t=\,$118\,yr. From left to right these are the adiabatic, the
rotating and the cooling PFD jets, respectively. Bottom
panels show a pole-on view.}
\end{wrapfigure}

PFD jets can be produced in pulsed power laboratory facilities and
help to understand the physics of astrophysical jets. The PFD ``lab jets''
are collimated by hoop stress, and their outer magnetic cavities
are collimated by external pressure.  The axial magnetic component
influences both the radial compression and the stability of the jets.
The outflows evolve to become corrugated, but still collimated, structures due to current-driven
instabilities. Thin conduction foil experiments produce episodic
jets and nested magnetic bubbles.  Observations of multiple lobes
in the radio galaxy B0925+420, may be consistent with such processes (Brocksopp et
al. \cite{epis}).

Our simulations show that  PFD jet beams are lighter, slower and
less stable than pre-collimated asymptotically hydrodynamic jets.
In practice the latter  could represent the asymptotic propagation
regimes  of  magneto-centrifugally launched jets, which are distinct
from PFD in that  PFD remain PFD out to much larger scales.  We
find that current-driven perturbations in PFD jets are amplified
by both cooling and rotation for the regimes studied: Shocks and
thermal pressure support are weakened by cooling, making the jets
more susceptible to kinking. Rotation amplifies the toroidal magnetic
field which also exacerbates the kink instability.  Our simulations
agree well with the models and experiments of Shibata \&
Uchida~(\cite{shibata}) and Lebedev et~al.~(\cite{leb5}), respectively.

\acknowledgements Financial support for this project was provided by the Space Telescope
Science Institute grants HST-AR-11251.01-A and HST-AR-12128.01-A;
by the National Science Foundation under award AST-0807363; by the
Department of Energy under award DE-SC0001063; by 
NSF  grant PHY0903797,  and by Cornell
University grant 41843-7012.


\begin{thebibliography}{}

\bibitem[2007]{b2007}Blackman E.G., 2007, Ap\&SS, 307, 7 
\bibitem[1982]{bland}Blandford, R. D., \& Payne, D. G., 1982, MNRAS, 199, 883
\bibitem[2007]{epis}
Brocksopp, C., Kaiser, C.~R., Schoenmakers, A.~P., \& de Bruyn, A.~G., 
2007, MNRAS, 382, 1019 
\bibitem[2011]{bear2}
Carroll-Nellenback, J.J., Frank, A., Shroyer, B., \& Ding, C., 2011 (in prep)
\bibitem[2007]{ciardi7}
Ciardi, A., Lebedev, S.~V., Frank, A., et al., 2007, Physics of Plasmas, 14, 056501
\bibitem[2009]{ciardi9}
Ciardi, A., Lebedev, S.~V., Frank, A., et al., 2009, ApJL, 691, L147
\bibitem[2009]{bear}
Cunningham A. J. et~al., 2009, ApJS, 182, 519
\bibitem[1972]{dm}
Dalgarno A., McCray R. A., 1972, ARA\&A, 10, 375
\bibitem[1999]{laser1} 
Farley, D.~R., Estabrook, K.~G., Glendinning, S.~G., et al., 1999, Physical Review 
Letters, 83, 1982
\bibitem[2002]{laser3} Foster, J.~M., Wilde, B.~H., Rosen, P.~A., et al., 2002, 
Physics of Plasmas, 9, 2251
\bibitem[2011]{we}
Huarte-Espinosa, M., Frank A., Blackman E. G., Lebedev S., Ciardi A., Hartigan, P., 2011, ApJ, (in prep)
\bibitem[1958]{kruskal} 
Kruskal, M.~D., Johnson, J.~L., Gottlieb, M.~B., Goldman, L.~M., 1958, Physics of Fluids, 1, 421 
\bibitem[2005]{leb5}
Lebedev, S. V., et al., 2005, MNRAS, 361, 97
\bibitem[2007]{pudritz}
Pudritz, R. E., et~al., 2007, Protostars and Planets V, 277
\bibitem[2004]{naka}
Nakamura, M., \& Meier, D. L., 2004, ApJ, 617, 123
\bibitem[1958]{shafranov} 
Shafranov, V.~D., 1958, Soviet Journal of Experimental and Theoretical Physics, 6, 545 
\bibitem[1986]{shibata} 
Shibata, K., \& Uchida, Y., 1986, PASJ, 38, 631
\bibitem[2000]{laser2} 
Shigemori, K., Kodama, R., Farley, D.~R., et al., 2000, PRE, 62, 8838 
\bibitem[2010]{suzuki} 
Suzuki-Vidal, F., Lebedev, S.~V., Bland, S.~N., et al., 2010, IEEE Transactions on Plasma Science, 38, 581
\end{thebibliography}
\end{document}